# Magnetization Switching in van der Waals Systems by Spin-Orbit Torque

Xin Lin[1,2], Lijun Zhu[1,2*]

1. *State Key Laboratory of Superlattices and Microstructures, Institute of Semiconductors, Chinese Academy of Sciences, Beijing 100083, China*
2. *College of Materials Science and Opto-Electronic Technology, University of Chinese Academy of Sciences, Beijing 100049, China*

*ljzhu@semi.ac.cn

**Abstract**: Electrical switching of magnetization via spin-orbit torque (SOT) is of great potential in fast, dense, energy-efficient nonvolatile magnetic memory and logic technologies. Recently, enormous efforts have been stimulated to investigate switching of perpendicular magnetization in van der Waals systems that have unique, strong tunability and spin-orbit coupling effect compared to conventional metals. In this review, we first give a brief, generalized introduction to the spin-orbit torque and van der Waals materials. We will then discuss the recent advances in magnetization switching by the spin current generated from van der Waals materials and summary the progress in the switching of Van der Waals magnetization by the spin current.

## 1. Introduction
### 1.1 Spin-orbit torque

Spin-orbit torques (SOTs) are a powerful tool to manipulate magnetization at the nanoscale for spintronic devices, such as magnetic random access memory (MRAM) and logic [1-5]. SOTs are exerted on a magnetization when angular momentum is transferred from spin accumulation or spin currents carried by a flow of electrons or magnons (Fig. 1). A spin current with spin polarization vector $\boldsymbol{\sigma}$, can exert two types of SOTs on a magnetization $\boldsymbol{M}$, i.e., a damping-like (DL) torque [$\boldsymbol{\tau}_{DL} \sim \boldsymbol{M} \times (\boldsymbol{M} \times \boldsymbol{\sigma})$] due to the absorption of the spin current component transverse to $\boldsymbol{M}$ and a field-like (FL) torque [$\boldsymbol{\tau}_{FL} \sim \boldsymbol{M} \times \boldsymbol{\sigma}$] due to the reflection of the spin current with some spin rotation. In the simple case of the spin-current generator/magnet bilayer, the efficiency of the damping-like SOT per unit bias current density, $\xi_{DL}^{j}$, can be estimated as [6]

$$\xi_{DL}^{j} \approx T_{int}\, \theta_{SH}\, \tau_M^{-1}/(\tau_M^{-1}+\tau_{so}^{-1}) \qquad (1)$$

where $\tau_M^{-1}/(\tau_M^{-1}+\tau_{so}^{-1})$ is the percentage of the spin current relaxed via the spin-magnetization exchange interaction (with spin relaxation rate $\tau_M^{-1}$) within the magnetic layer and is less than 1 in presence of non-negligible spin relaxation via the spin-orbit scattering (with spin relaxation rate $\tau_M^{-1}$) [6], $T_{int}$ is the interfacial spin transparency which determines what fraction of the spin current enters the magnet (less than 1 in presence of spin backflow [7-11] and spin memory loss [12-15]), and $\theta_{SH}$ is the charge-to-spin conversion efficiency of the spin current generator (e.g., the spin Hall ratio in the case of spin Hall materials [2-4]). The quantitative understanding of the efficiency of the field-like torque, $\xi_{FL}^{j}$, remains an open question.

The same SOT physics can be expressed in terms of effective SOT fields: a damping-like effective SOT field ($H_{DL}$) parallel to $\boldsymbol{M} \times \boldsymbol{\sigma}$ and a field-like effective SOT field ($H_{FL}$) parallel to $\boldsymbol{\sigma}$. The magnitudes of the damping-like and field-like SOT fields correlate to their SOT efficiencies per unit bias current density via

$$H_{DL} = (\hbar/2e)\, j\xi_{DL}^{j} M_s^{-1} t^{-1} \qquad (2)$$

$$H_{FL} = (\hbar/2e)\, j\xi_{FL}^{j} M_s^{-1} t^{-1} \qquad (3)$$

where $e$ is the elementary charge, $\hbar$ reduced Plank's constant, $t$ the magnetic layer thickness, $M_s$ the saturation magnetization of the magnetic layer, and $j$ the charge current density in the spin current generating layer.

The damping-like SOT is technologically more important because it can excite dynamics and switching of magnetization (even for low currents for which $H_{DL}$ is much less than the anisotropy field of a perpendicular magnetization). Field-like SOTs by themselves can destabilize magnets only if $H_{FL}$ is greater than the anisotropy field, but they can still strongly affect the dynamics in combination with a damping-like SOT [16-18].

The generation of spin currents is central to the SOT phenomena. The spin polarization vector $\boldsymbol{\sigma}$ can have longitudinal, transverse, and perpendicular components, i.e., $\sigma_x$, $\sigma_y$, and $\sigma_z$. Transversely polarized spin current can be generated by a longitudinal charge current flow in either magnetic or non-magnetic materials via a variety of possible spin-orbit-coupling (SOC) effects. The latter includes the bulk spin Hall effect (SHE) [19-22], topological surface states [23,24], interfacial SOC effects [16,25-28], orbit-spin conversion [29], the anomalous Hall effect [30,31], the planar Hall effect [32,33], the magnetic SHE [34-36], Dresselhaus effect [37], Dirac nodal lines [38,39], etc. The bulk SHE has been widely observed in thin-film heavy metal (HM) [40-42], Bi-Sb [43], $Bi_xTe_{1-x}$ [44], CoPt [45], FePt [46], $Fe_xTb_{1-x}$ [47], and Co-Ni-B [48].

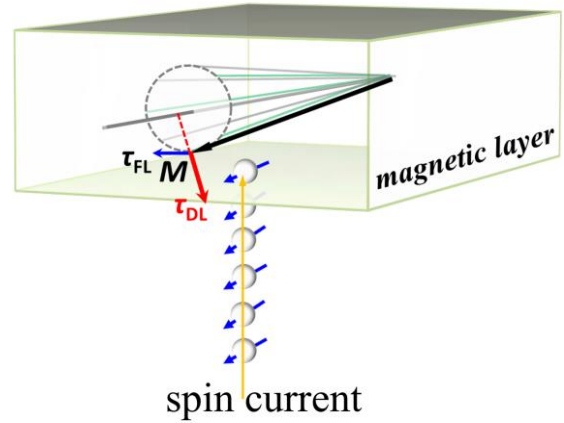

Fig. 1 Schematic of damping-like and field-like spin-orbit torques exerted on a magnetic layer by an incident spin current.

Since the transverse spins, in principle, cannot switch a uniform perpendicular magnetization without the assistance of an in-plane longitudinal magnetic field either via coherent rotation of macrospin or via domain wall depinning [2-4,49], generation of perpendicular and longitudinal spins [50-53] are of great interests. While perpendicular and longitudinal spins are not allowed in nonmagnetic materials that are cubic crystal or polycrystalline/amorphous, additional crystal or magnetic symmetry breaking can be introduced to make perpendicular and longitudinal spins permissive.

Generation of perpendicular spins has been argued from low-symmetry crystals (e.g., WTe$_2$ [50], MoTe$_2$ [54,55], and CuPt [56]), non-collinear antiferromagnetic crystals with magnetic asymmetry (e.g., IrMn$_3$ [57], Mn$_3$GaN [58,59], Mn$_3$Sn [34]), some collinear antiferromagnets with spin conversions (e.g., Mn$_2$Au [60,61] and RuO$_2$ [62-64]), and also some magnetic interfaces [65]. Longitudinal spins might be generated by low-symmetry crystals (*e.g.*, MoTe$_2$ [54], (Ga, Mn)As [66], NiMnSb [67], and Fe/GaAs [68]) or by a non-zero perpendicular magnetization [69,70].

**1.2 van der Waals materials**

So far, the most widely used spin-source materials are heavy metals with strong spin Hall effect (e.g., Pt with a giant spin Hall conductivity of 1.6×10$^6$ ($\hbar$/2e) Ω$^{-1}$ m$^{-1}$ in the clean limit) [71], while the 3$d$ ferromagnets (e.g., Co, Fe, Ni$_{81}$Fe$_{19}$, CoFeB) and ferrimagnets (e.g., FeTb, CoTb, and GdFeCo) are well studied as the spin-detectors. However, it is believed that the energy efficiency of SOT devices may be improved by developing new materials and new mechanisms that generate spin currents.

Van der Waals materials have attracted enormous attention in the field of material science and condensed matter physics since the discovery of single monolayer graphene in ref. [72] and is increasingly investigated in the field of spintronics. This is particularly due to the diversity of materials, the flexibility of preparation (e.g., by mechanical exfoliation), and strong tunability by interface effects. In the field of spintronics, the van der Waals materials are also interesting for intriguing SOC effects. For example, the non-magnetic van der Waals materials of transition metal dichalcogenides (TMDs) and topological insulators (TIs)[73,74] exhibit a strong ability in generating transversely polarized spin current via the spin Hall effect and/or topological surface states [75-77], in some cases also in generating currents of perpendicular and longitudinal spins due to low crystal symmetry [50,51]. On other hand, van der Waals materials that are magnetic are also interesting for spin-orbitronics due to their highly tunable magnetism and low magnetization at small thicknesses [78-83]. As indicated by Eq. (2) and Eq. (3), the damping-like and field-like effective SOT fields exerted on the magnetic layer by a given spin current scale inversely with the thickness and magnetization of the magnetic layer.

This review is intended to focus on highlighting the recent advances of magnetization switching in van der Waals systems by spin-orbit torque, including switching of conventional magnetic metals by spin current from non-magnetic van der Waals TMDs and TIs and switching of van der Waals magnets by incident spin currents.

**2. Magnetization switching by spin current from Transition Metal Dichalcogenide semimetals**

TMDs typically consist of transition metals (e.g., Mo, W, Pt, Ta, Zr) and chalcogenide elements (e.g., S, Te, Se). From point of view of SOT applications, it is most interesting to develop the TMD semi-metals with relatively high conductivity, strong spin-orbit coupling, high $\theta_{SH}$, and reduced structural symmetry. However, TMD semiconductors (e.g., MoS$_2$ [84,85], WS$_2$ [86,87], WSe$_2$ [85]) are merely studied in spin-orbitronics due to their very poor conductivity that is detrimental to the energy TMD metals (e.g., TaS$_2$ [88], NbSe$_2$ [89]) are highly conductive but typically not efficient in generating spin currents. Therefore, below we mainly discuss the progress in spin-orbit torque studies of TMD semimetals that are interesting for SOT applications(e.g., WTe$_2$ [50,90,91], PtTe$_2$ [92], MoTe$_2$ [51,93], and ZrTe$_2$ [94]).

**2.1 Exfoliated Transition Metal Dichalcogenide semimetals**

The pioneering studies of TMD semimetals in the field of spintronics [50] mainly focused on the characterization of damping-like and field-like SOTs of transverse, perpendicular, and longitudinal spins generated in bilayers of mechanically exfoliated TMD semimetals and 3$d$ ferromagnets (e.g., Ni$_{81}$Fe$_{19}$). These studies have opened a new subject field that unitizes van der Waals materials for the possible generation of transverse, perpendicular, and longitudinal spin polarizations.

WTe$_2$ is a semimetal with a low inversion symmetry along the *a* axis of the crystal (the space group *Pmn*2$_1$) [Fig. 2(a)] and the Weyl points at the crossing of the oblique conduction and the valence bands only at low temperatures (typically below 100 K [95]). In a WTe$_2$/ferromagnet bilayer, the screw-axis and glide-plane symmetries of this space group are broken at the interface, so that WTe$_2$/ferromagnet bilayers have only one symmetry, a mirror symmetry relative to the *bc* plane (depicted in Fig. 2(a)). There is no mirror symmetry in the *ac* plane, and therefore no 180° rotational symmetry about the *c* axis (perpendicular to the sample plane). MacNeill *et al.* [50] first observed in mechanically exfoliated WTe$_2$/Ni$_{81}$Fe$_{19}$ bilayers damping-like (≈ 8×10$^3$ ($\hbar$/2e) Ω$^{-1}$ m$^{-1}$) and field-like spin-orbit torques of transverse spins at room temperature as well as the exotic damping-like SOT of perpendicular spins (≈ 3.6×10$^3$ ($\hbar$/2e) Ω$^{-1}$ m$^{-1}$). The damping-like SOT of perpendicular spins manifests as an additional sin2$\varphi$ term in the antisymmetric signal of spin-torque ferromagnetic resonance (ST-FMR)[Fig. 2(b)] and was attributed to the symmetry breaking at the interface of the WTe$_2$ crystal. The damping-like SOT of perpendicular spins is found to maximize when current is applied along the low-crystal-symmetry *a* axis and vanishes when current is applied along the high-crystal-symmetry *b* axis [50]. This is in contrast to the torques of the transverse spins that are independent of the crystal orientation. The damping-like SOT of perpendicular spins in WTe$_2$/Ni$_{81}$Fe$_{19}$ was also found to vary little with the WTe$_2$ thickness, which was suggested as an indication that the spin current is mainly generated near the interface of the WTe$_2$ [90,91,96]. Xie *et al.* [52] reported that in-plane direct current along the *a* axis of WTe$_2$ can induce partial switching of magnetization in absence of an external magnetic field [Fig. 2(c)] and shift of the anomalous Hall resistance loop in SrRuO$_3$/exfoliated WTe$_2$ bilayers, which was speculated as an indication of damping-like torque of perpendicular spins on the perpendicular magnetization (macrospin). However, there can be longitudinal and perpendicular Oersted fields due to current spreading in the WTe$_2$ layer [90,91,96], which can also induce "field-free" switching of perpendicular magnetization and anomalous Hall loop shifts via adding to or subtracting from the domain wall depinning field (coercivity). The SOTs of the WTe$_2$/Ni$_{81}$Fe$_{19}$ have also been reported to switch the Ni$_{81}$Fe$_{19}$ layer with weak in-plane magnetic anisotropy at a current density of $\approx 3\times10^5$ A cm$^{-2}$ [91]. An in-plane current along the $a$ axis of WTe$_2$ has also been reported to enable partial switching of perpendicular magnetization in WTe$_2$/Fe$_{2.78}$GeTe$_2$ without an external magnetic field [53].

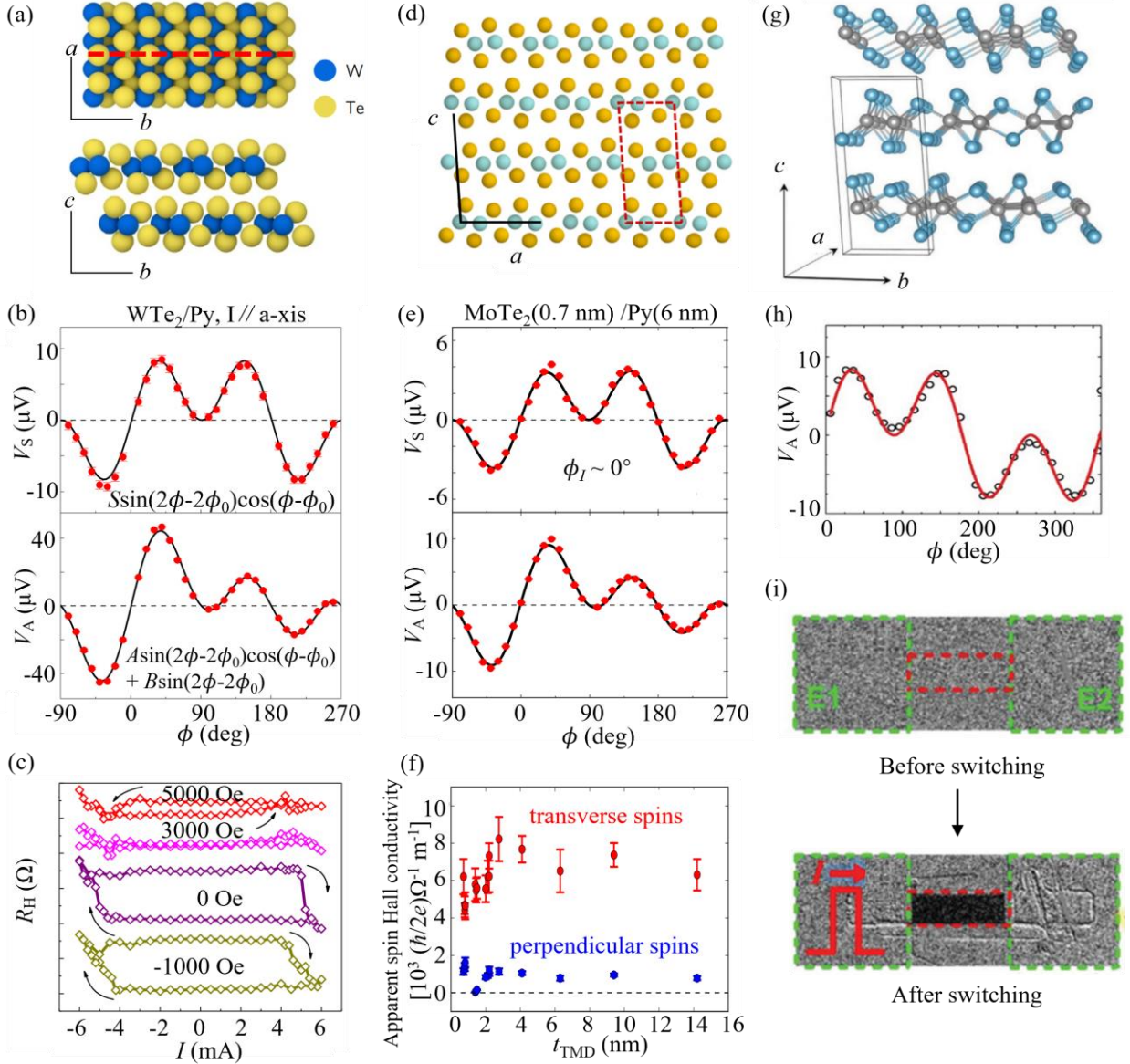

FIG. 2. (a) Crystal structure near the surface of WTe$_2$, displaying a mirror symmetry relative to the $bc$ plane but not to the $ac$ plane. (b) Symmetric ($V_S$) and antisymmetric ($V_A$) components of ST-FMR signal for WTe$_2$ (5.5)/Py (6) device as a function of the angle of the in-plane magnetic field [50]. Reprinted with permission from MacNeill et al., Nat. Phys. 13, 300 (2017). (c) Current-induced magnetization switching of WTe$_2$(15)/SrRuO$_3$ when the current is along the low-symmetry $a$ axis where the magnetization can be switched without an external magnetic field [52]. Reprinted with permission from Xie et al., APL Mater. 9, 051114 (2021). (d) Structure of the MoTe$_2$ crystal in the monoclinic ($\beta$ or 1T′) phase depicted in the $a$-$c$ plane for which the mirror plane is within the page and the Mo chains run into the page. (e) Symmetric and antisymmetric ST-FMR resonance components for the MoTe$_2$(0.7)/Py(6) device with a current applied perpendicular to the MoTe$_2$ mirror plane as a function of the orientation of the in-plane magnetic field. (f) The conductivities of damping-like torque of perpendicular spins (blue) and transverse spins (red) as a function of the MoTe$_2$ thickness for devices with current aligned perpendicular to the MoTe$_2$ mirror plane [51]. Reprinted with permission from Stiehl et al., Phys. Rev. B 100, 184402 (2019). (g) Crystal structure of the monoclinic 1T′ phase of MoTe$_2$. (h) Antisymmetric ST-FMR components for MoTe$_2$(83.1)/Py(6) as a function of the orientation of the in-plane magnetic field. (i) MOKE images implying switching of Py by current [93]. Reprinted with permission from Liang et al., Adv. Mater. 32, 2002799 (2020).

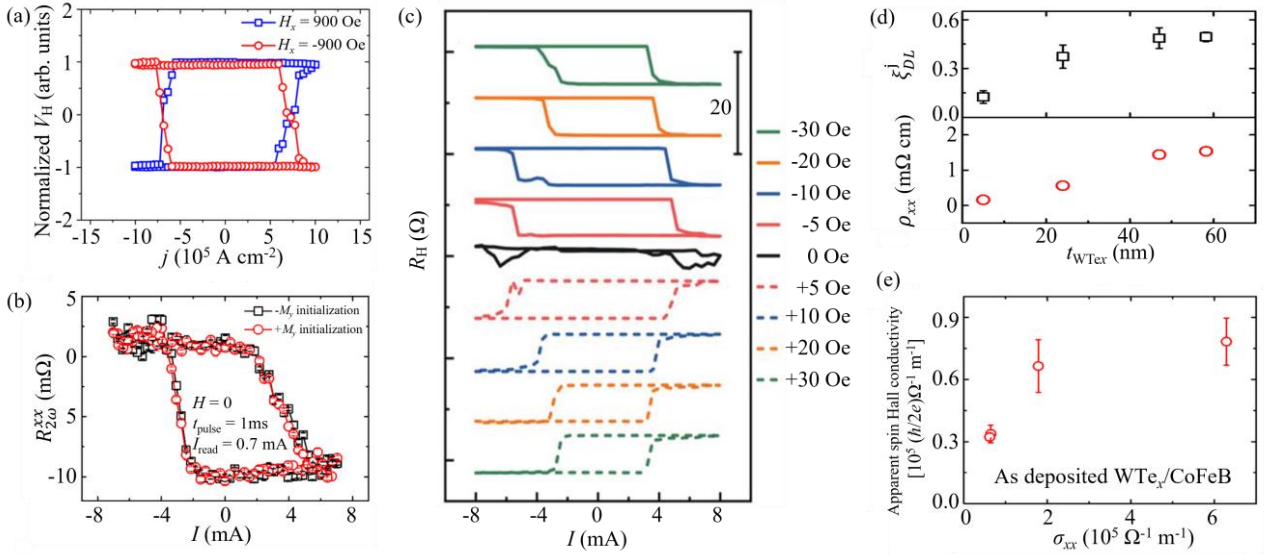

FIG. 3. (a) Current-induced magnetization switching in sputter-deposited WTe$_2$(10)/CoTb(6)/Ta(2) Hall bar ($H_x= \pm 900$ Oe) [97]. Reprinted with permission from Peng et al., ACS Appl. Mater. Interfaces 13, 15950 (2021). (b) Second harmonic longitudinal resistance ($R_{2\omega}^{xx}$) of WTe$_x$(5)/Mo(2)/CoFeB(1) measured as a function of pulse current amplitude $I_{pulse}$ under zero external field [98]. (c) Current-induced switching loops of WTe$_x$(5)/Ti(2)/CoFeB(1.5) Hall bar under different in-plane magnetic fields at 200 K [99]. Reprinted with permission from Xie et al., Appl. Phys. Lett. 118, 042401 (2021). (d) Dependences on the WTe$_x$ thickness of damping-like SOT efficiency ($\xi_{DL}^j$) and the WTe$_x$ resistivity ($\rho_{xx}$) for WTe$_2$/CoFeB. (e) Apparent spin Hall conductivity as a function of the longitudinal conductivity for WTe$_2$/CoFeB [98]. Data in (b), (d), and (e) are reprinted with permission from Li et al., Matter 4, 1639 (2021).

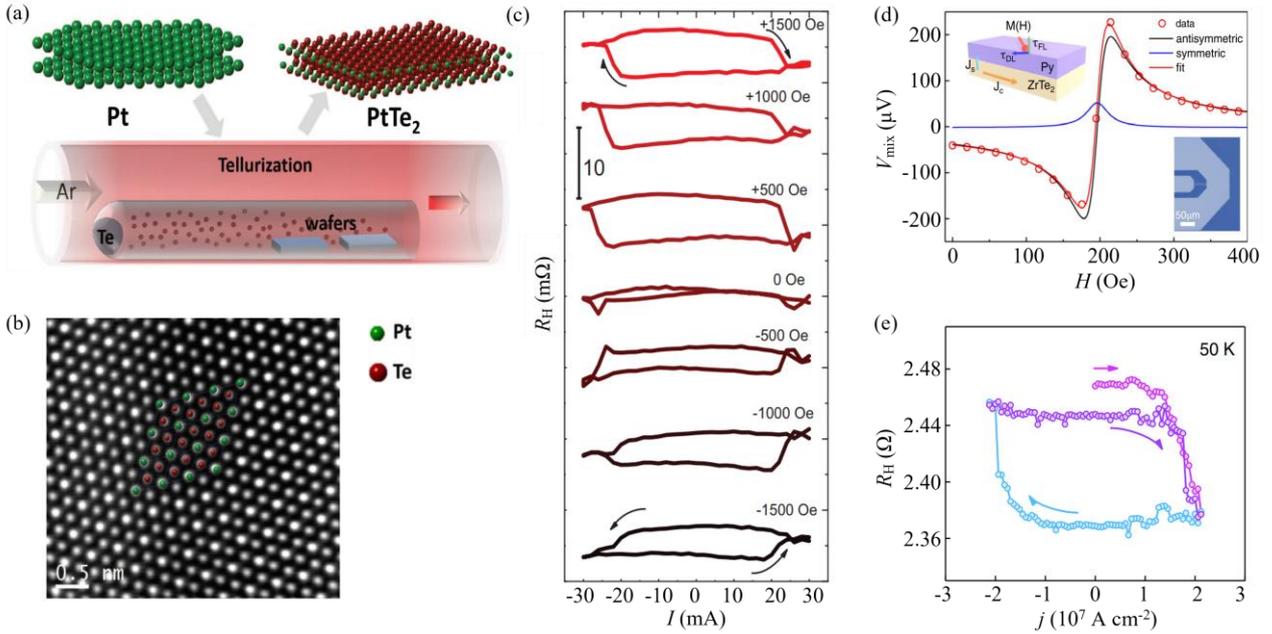

FIG. 4. (a) Schematic of the CVD growth process for PtTe$_2$. (b) High-resolution transmission electron microscopy image of a 5 nm PtTe$_2$ thin film. (c) Current-induced magnetization switching in the PtTe$_2$(5)/Au(2.5)/CoTb(6) Hall bar under different in-plane fields [92]. Reprinted with permission from Xu et al., Adv. Mater. 32, 2000513 (2020). (d) Spin torque ferromagnetic resonance spectrum of a ZrTe$_2$/Py bilayer at room temperature. (e) Current-induced magnetization switching in ZrTe$_2$(8 u.c.)/CrTe$_2$(3 u.c.) Hall bar under a 700 Oe in-plane field at 50 K [94]. Reprinted with permission from Ou et al., Nat. Commun. 13, 2972 (2022).

$\beta$-MoTe$_2$ is a semimetal that retains inversion symmetry in bulk but has a low-symmetry interface (the group space is $Pmn2_1$ in bulk but $Pm11$ in few-layer structures [Fig. 2(d)]). Stiehl et al. [51] observed damping-like SOT of both transverse spins ($\approx 8\times10^3$ ($h/2e$) $\Omega^{-1}$ m$^{-1}$) and perpendicular spins ($\approx 1\times10^3$ ($h/2e$) $\Omega^{-1}$ m$^{-1}$) in mechanically exfoliated $\beta$-MoTe$_2$/Ni$_{81}$Fe$_{19}$ bilayers [Fig. 2(e)]. This torque of perpendicular spins is one-third strong than that of WTe$_2$/Ni$_{81}$Fe$_{19}$ and was attributed to perpendicularly polarized spin current from the surface of the low-symmetry $\beta$-MoTe$_2$ [Fig. 2(f)]. This appears to suggest that the breaking of bulk inversion symmetry is not an essential requirement for producing perpendicular spins. However, 1T'-MoTe$_2$ [Fig. 2(g)] was reported to generate no damping-like SOT of perpendicular spins in contact with $Ni_{81}Fe_{19}$ [Fig. 2(h)][93]. Instead, 1T′-$MoTe_2$ only generates a nonzero damping-like SOT of transverse spins that switches the in-plane magnetized $Ni_{81}Fe_{19}$ layer at a current density of $6.7×10^5$ A $cm^{-2}$ [Fig. 2(i)]. $NbSe_2$ with resistivity anisotropy was reported to generate a perpendicular Oersted field but no perpendicular or longitudinal spins when interfaced with $Ni_{81}Fe_{19}$ [89]. The damping-like toque of transverse spins in mechanically exfoliated $NbSe_2$/$Ni_{81}Fe_{19}$ is very weak and corresponds to a spin Hall conductivity of ≈ $10^3$ ($ℏ/2e$) $Ω^{-1}$ $m^{-1}$ [89].

Here it is important to note that, while the presence of perpendicular spins has been widely concluded in the literature from a sin2$φ$-dependent contribution in *symmetric* spin-torque ferromagnetic resonance signal of in-plane magnetization ($φ$ is the angle of the external magnetic field relative to the current), or a $φ$-independent but field-dependent contribution in the second harmonic Hall voltage of in-plane magnetization, or field-free switching, none of the three characteristics can simply "signify" the presence of a flow of perpendicular spins. This is because non-uniform current effects that can generally exist and generates out-of-plane Oersted field in nominally uniform, symmetric Hall bars and ST-FMR strips [90,91,96,100] also exhibit all three characteristics. As demonstrated by Liu and Zhu [100], these characteristics can be considerable especially when the devices have strong current spreading, e.g., in presence of non-symmetric electric contacts.

**2.2 Large-area Transition Metal Dichalcogenides**

So far, most TMD studies have been based on mechanical exfoliation, which is unsuitable for the mass production of spintronic applications. Recently, efforts have been made in large-area growth of thin-film TMDs towards the goal of SOT applications [97,98]. For example, sputter-deposited $WTe_x$ has also developed to drive low-current-density switching of CoTb ($j_c$ ≈ $7.05×10^5$ A $cm^{-2}$ under in-plane assisting field of 900 Oe) [Fig. 3(a)] [97] and in $WTe_x$/Mo/CoFeB ($j_c$ ≈ $7×10^6$ A $cm^{-2}$, under no in-plane assisting field, Fig. 3(b))[99] and in $WTe_x$/Ti/CoFeB ($j_c$ ≈ $2.0×10^6$ A $cm^{-2}$ under in-plane assisting field of ± 30 Oe, Fig. 3(c))[98]. It has become a consensus that the spin-orbit torque in these sputter-deposited $WTe_x$/FM samples arises from the bulk spin Hall effect of the $WTe_x$ [97,98]. As indicated in Figs. 3(d) and 3(e), the measured spin-orbit torque efficiency increases but the apparent spin Hall conductivity decreases as the resistivity increases in the dirty limit [41] due to increasing layer thickness.

Large-area $PtTe_2$ films with relatively high electrical conductivity (≈ $3.3×10^6$ $Ω^{-1}$ $m^{-1}$ at room temperature) and spin Hall conductivity ($2×10^5$ $ℏ/2e$ $Ω^{-1}m^{-1}$) have also been reported by annealing a Pt thin film in tellurium vapor at ≈ 460 °C [Figs. 4(a) and 4(b)][92]. $PtTe_2$ is predicted to be a type-II Dirac semimetal with spin-polarized surface states. However, there is no indication of the generation of torques of out-of-plane spins. Partial switching of magnetization by in-plane current has also been reported in a $PtTe_2$(10)/Au(2.5)/CoTb(6) Hall bar ($j_c$ ≈ $9.9×10^6$ A $cm^{-2}$ under in-plane assisting field of 2 kOe) [Fig. 4(c)].

Growth of $ZrTe_2$ by molecular beam epitaxy (MBE) has also been reported. A ST-FMR study has measured a small damping-like torque of transverse spins for MBE-grown $ZrTe_2$/$Ni_{81}Fe_{19}$ bilayer at room temperature [Fig. 4(d)][94]. This is consistent with the theoretical prediction that $ZrTe_2$ is a Dirac semimetal with massless Dirac fermions in its band dispersion [101] but vanishing spin Hall conductivity. Even so, a $ZrTe_2$(8 u.c.)/$CrTe_2$(3 u.c.) bilayer has been reported to be partially switched at 50 K by an in-plane current of $1.8×10^7$ A $cm^{-2}$ in density under an in-plane assisting field of 700 Oe [Fig. 4(e)].

**3. Magnetization switching by spin current from Bi-based topological insulators**

Another kind of layered strong-SOC material is Bi-based topological insulators [102-104]. As displayed in Fig. 5 (a), TIs are insulating in the bulk but conducting at the surface. The initial interest of TIs for spin-orbit torque studies is the topological surface states (Fig. 5 (b)). In the wavevector space, the spin and momentum of electrons are one-to-one locked to each other at the Fermi level. With a flow of charge current, the shift in the electron distribution in the wavevector space induces non-equilibrium spin accumulation (Fig. 5 (c)).

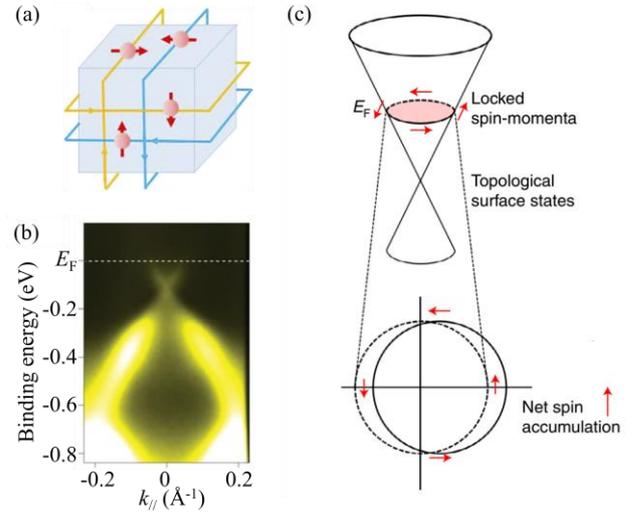

Fig. 5. Topological surface states and spin-accumulation in topological insulators. (a) Real-space picture of the conducting surface states in an ideal topological insulator [103]. Reprinted with permission from Han and Liu, APL Mater. 9, 060901 (2021). (b) Angle-resolved photoemission spectrum that indicates the bulk and surface bands of a six-quintuple-layer-thick $Bi_2Se_3$ film [102]. Reprinted with permission from Zhang *et al*., Nat. Phys. 6, 584 (2010). Copyright 2010 Springer Nature Limited. (c) Current-induced spin accumulation in a topological insulator [104]. The arrows denote the directions of spin magnetic moments, which are opposite to the corresponding spin angular momenta. Reprinted with permission from He *et al*., Nat. Mater. 21, 15–23 (2022).

**3.1 MBE-grown and exfoliated Topological insulators**

Topological insulators were first introduced in the field of spin-orbit torque in 2014. From ST-FMR measurement, Mellnik *et al*. measured a giant damping-like spin-orbit torque efficiency ($\xi_{DL}^j$ = 3.5) at room temperature in $Bi_2Se_3$/Py bilayers grown by MBE [Fig. 6(a)][23]. In the same year, Fan *et al*. reported from harmonic Hall measurement a damping-like torque efficiency of 425 and the spin–orbit torque switching in the

$(Bi_{0.5}Sb_{0.5})_2Te_3/(Cr_{0.08}Bi_{0.54}Sb_{0.38})_2Te_3$ bilayers [24] at 1.9 K [Fig. 6(b)].

As shown in Fig. 7, room-temperature magnetization switching by spin current from TIs (e.g., $Bi_2Se_3$, $Bi_2Te_3$, and BiSb) has been demonstrated in Hall-bar samples [99,105]. Han et al. [76,106] first reported magnetization switching in Hall bars of $Bi_2Se_3(7.4)/Co_{0.77}Tb_{0.23}(4.6)$ bilayer ($H_x$ = 1000 Oe, $H_c \approx$ 200 Oe, $J_c \approx 2.8\times10^6$ A cm$^{-2}$, switching ratio=85%)[Fig. 7(a)]. The damping-like SOT efficiency was determined to be 0.16 ± 0.02 for the $Bi_2Se_3/Co_{0.77}Tb_{0.23}$. Similar results have been also reported by Wu et al. [77] in $Bi_2Se_3/Gd_x(FeCo)_{1-x}$ Hall bars ($\xi_{DL}^j$ = 0.13, $J_c \approx 2.2\times10^6$ A cm$^{-2}$)[Fig. 7(b)]. These values of spin-orbit torque efficiency are significantly low compared to those from $Bi_2Se_3$/Py samples, which may be understood partly by the increased spin current relaxation via spin-orbit scattering in the ferrimagnets [6]. Khang et al. have reported a spin-orbit torque efficiency of 52 (as determined from a coercivity change measurement) and resistivity of 400 μΩ cm for MBE-grown $Bi_{1-x}Sb_x$ [107]. Switching of MBE-grown fully epitaxial $Mn_{0.45}Ga_{0.55}/Bi_{0.9}Sb_{0.1}$ has also been demonstrated at a current density of $1.1\times10^6$ A cm$^{-2}$ ($H_x$ = 3.5 kOe) in [Fig. 7(c)]. Non-epitaxial BiSb films (10 – 20 nm) grown by MBE were also reported to have a high spin-orbit torque efficiency of up to 3.2 and to enable magnetization switching at a current density of $2.2\times10^6$ A cm$^{-2}$. There have also been reports of SOT switching of exfoliated van-der-Waals magnets at low temperatures, such as in $Fe_3GeTe_2$ [108,109] and $Cr_2Ge_2Te_6$ [110,111], by the spin current from TIs. Liu et al. [112] reported a strongly temperature-dependent damping-like torque efficiency of up to 70 from a field-dependent harmonic Hall response measurement[Fig. 8(a)], and current switching of MBE-grown $Bi_2Te_3$/MnTe Hall bar at a critical current density of down to $6.6\times10^6$ A cm$^{-2}$ ($H_x$ = ± 400 Oe, $T$ = 90 K) [Fig. 8(b)].

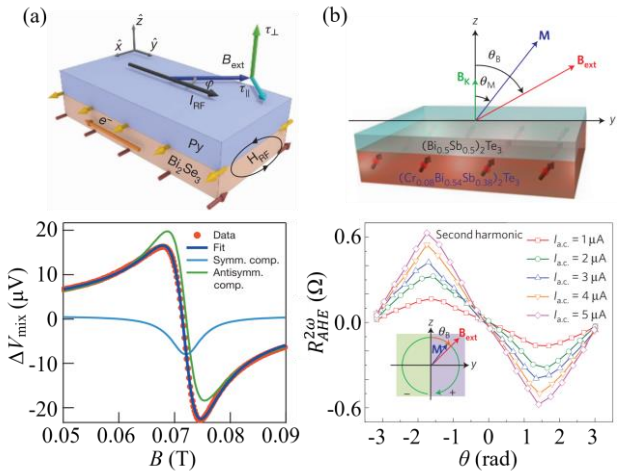

FIG. 6. (a) Spin torque ferromagnetic resonance spectrum for $Bi_2Se_3(8)/Py(16)$ bilayer at room temperature [23]. Reprinted with permission from Mellnik et al., Nature 511, 449 (2014). (b) Second harmonic Hall resistance for $(Bi_{0.5}Sb_{0.5})_2Te_3$(3 QL)/$(Cr_{0.08}Bi_{0.54}Sb_{0.38})_2Te_3$(6 QL) bilayer as a function of the in-plane field angle for different applied a.c. current [24]. Reprinted with permission from Fan et al., Nat. Mater. 13, 699 (2014).

## 3.2 Sputter-deposited Topological Insulators

Since exfoliation and molecular-beam epitaxy are less realistic methods for the preparation of large-area TI thin films for practical SOT devices, sputtering has been introduced to grow amorphous or polycrystalline "topological insulators". The first report of sputter-deposited "topological insulators" is $Bi_xSe_{(1-x)}$ [113] with relatively high electrical conductivity ($0.78\times10^5$ Ω$^{-1}$m$^{-1}$ for 4 nm thickness). Such sputter-deposited $Bi_xSe_{(1-x)}$ exhibits a very high damping-like spin-torque efficiency of 18 and enabled magnetization switching in a $Bi_xSe_{(1-x)}$(4)/Ta(0.5)/CoFeB(0.6)/Gd(1.2)/CoFeB(1.1) at a low current density of ≈ $4.3\times10^5$ A cm$^{-2}$ [Fig. 8(c)]. Wu et al. [105] also reported room-temperature witching of $Bi_2Te_3$/Ti/CoFeB at a current density of $2.4\times10^6$A cm$^{-2}$ ($H_x$ = 100 Oe). In the Hall bar of sputter-deposited PMA $Bi_2Te_3(8)$/CoTb(6) bilayer, current-induced magnetization switching was reported at a low critical current density of $9.7\times10^5$ A cm$^{-2}$ [Fig. 8(d)]. Sputter-deposited BiSb films (10 nm) were reported to provide a spin-torque efficiency of 1.2 and to drive switching of CoTb at $4\times10^5$ A/cm$^2$ [106].

## 3.3 Practical impact

As we have discussed above, some TIs and their sputter-deposited counterparts are reported to have much higher damping-like torque efficiency than heavy metals. Meanwhile, the sputter-deposited TIs are typically several times more resistive than the MBE-grown ones since disordered films typically have stronger electron scattering than crystalline films. However, for practical SOT applications, the spin-source materials are required to have low resistivity and large damping-like spin-orbit torque efficiency. Despite their amazingly high damping-like spin-orbit torque efficiency, most TIs are highly resistive (> $1\times10^3$ μΩ cm), much more resistive than ferromagnetic metals in metallic spintronic devices (e.g., 110 μΩ cm for CoFeB). Current shunting into the adjacent metallic layers would be considerably more than that flows within the topological insulator layer, resulting in increases in the total switching current and power consumption of devices.

## 3.4 Mechanism of the spin current generation

Despite the debate, the two main mechanisms via which the TIs and their disordered counterparts generate spin current or spin accumulation are the spin Hall effect and the surface states. As suggested by Khang et al. [82], Chi et al. [43], Tian et al. [44], the bulk spin Hall effect is the dominant source of the spin current for the spin-orbit torque in $Bi_{0.9}Sb_{0.1}$, $Bi_{0.53}Sb_{0.47}$, and $Bi_xTe_{1-x}$. As shown in Figs. 9(a) and 9(b), in disordered $Bi_{0.53}Sb_{0.47}$ the apparent spin Hall conductivity increases non-linearly with increasing layer thickness, which is a typical spin diffusion behavior and in good consistent with a bulk spin Hall effect being the mechanism of the spin current generation. In contrast, the surface states of the TIs have been suggested to be the main spin current source in MBE-grown $(Bi_{1-x}Sb_x)_2Te_3$ [105] and $Bi_2Te_3$[112]. This suggestion is consistent with the strong dependence of the damping-like spin-orbit torque on the composition [105], the temperature [112], and thus the location of the Fermi level relative to the Dirac point [Figs. 9(c) - 9(e)][105,114]. In addition, DC et al. suggested that the quantum confinement effect of small grains should account for the high spin-torque efficiency in the sputter-deposited $Bi_xSe_{(1-x)}$ [113].

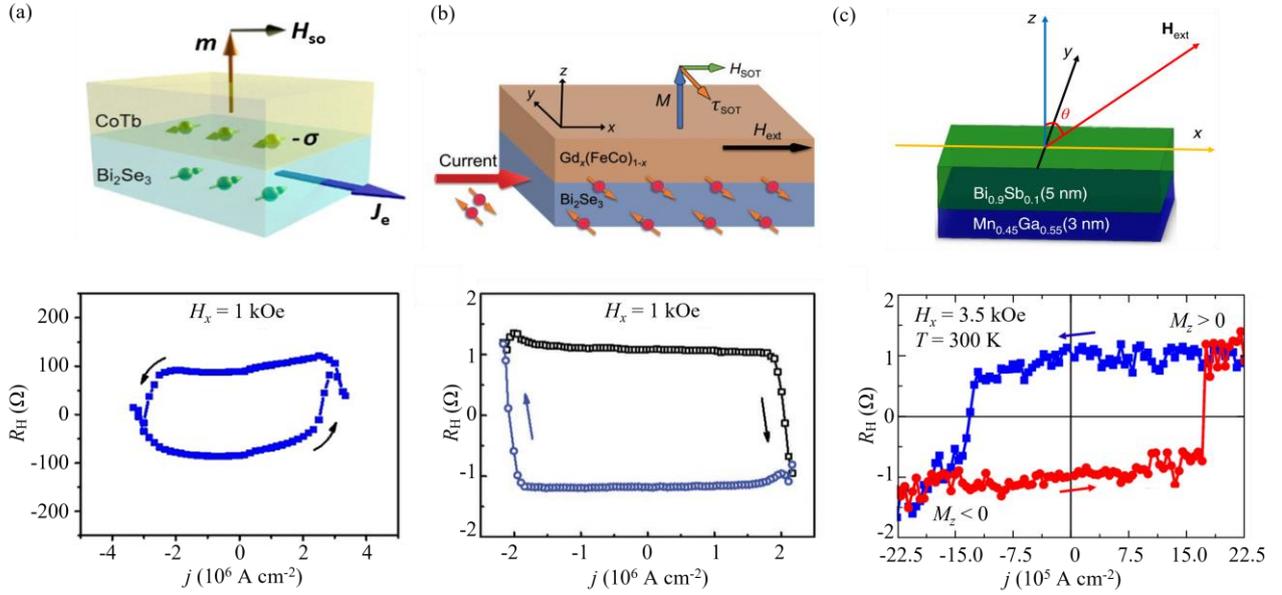

FIG. 7. Current-induced magnetization switching at room temperature in (a) $Bi_2Se_3(7.4)/CoTb(4.6)$ bilayer (the in-plane magnetic field is 1000 Oe) [76], (b) $Bi_2Se_3(6)/Gd_x(FeCo)_{1-x}(15)$ bilayer (an in-plane magnetic field is 1000 Oe) [77], and (c) $Mn_{0.45}Ga_{0.55}(3)/Bi_{0.9}Sb_{0.1}(5)$ (3.5 kOe) [115]. Data in (a) is reprinted with permission from Han *et al.*, Phys. Rev. Lett. 119, 077702 (2017). Data in (b) is reprinted with permission from Wu *et al.*, Adv. Mater. 31, 1901681 (2019); Data in (c) is reprinted with permission from Khang *et al.*, Nat. Mater. 17, 808 (2018).

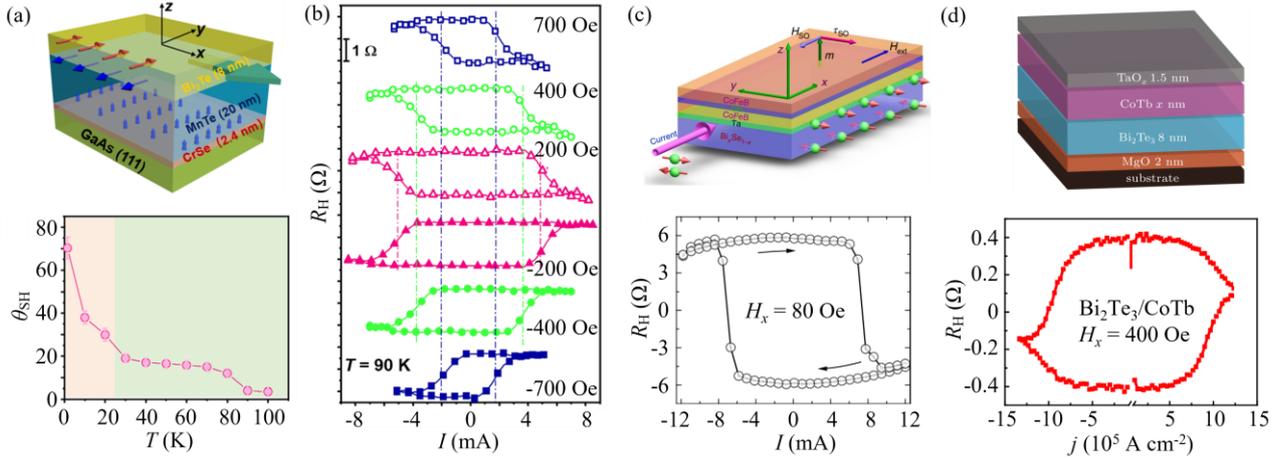

FIG. 8. (a) Variation of the spin Hall ratio of $Bi_2Te_3$ with temperature. (b) Current-induced magnetization switching of $Bi_2Te_3(8)/MnTe(20)$ at 90 K under different in-plane magnetic fields [112]. Reprinted with permission from Liu *et al.*, Appl. Phys. Lett. 118, 112406 (2021). (c) $Bi_xSe_{(1-x)}(4)/Ta(0.5)/CoFeB(0.6)/Gd(1.2)/CoFeB(1.1)$ (the in-plane magnetic field is 80 Oe)[113], reprinted with permission from DC *et al.*, Nat. Mater. 17, 800 (2018). (d) Current switching of sputter-deposited $Bi_2Te_3(8)/CoTb(6)$ under an in-plane magnetic field of 400 Oe at room temperature [116]. Reprinted with permission from Zheng *et al.*, Chin. Phys. B 29, 078505 (2020).

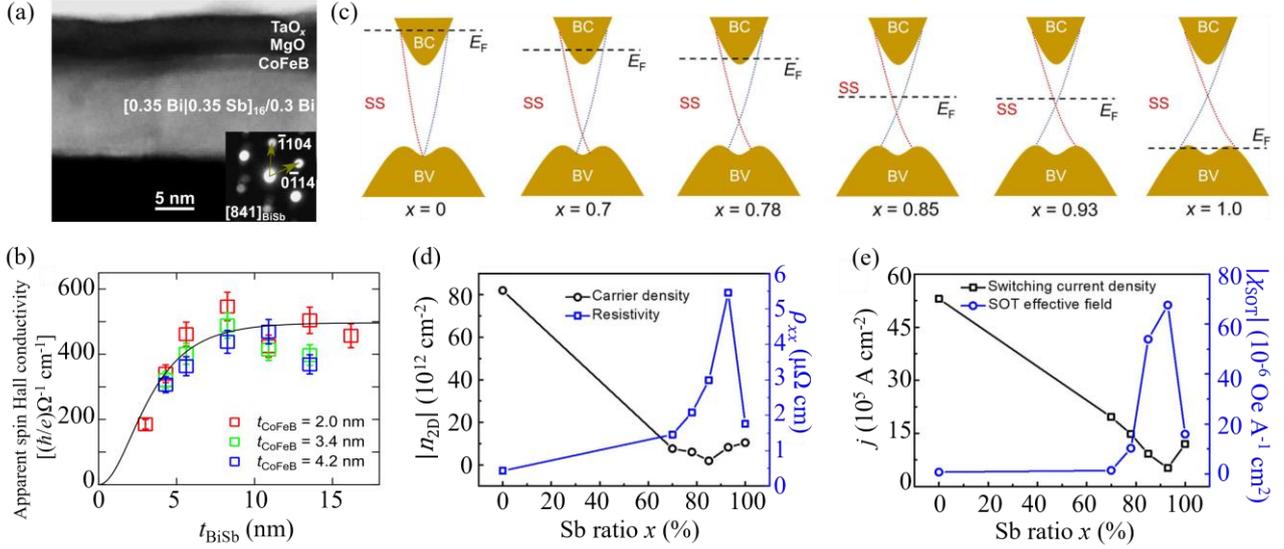

FIG. 9. (a) Scanning transmission electron microscopy image of the 0.5 Ta/[0.35 Bi|0.35 Sb]$_N$/0.3 Bi/2 CoFeB/2 MgO/1 Ta structure with N = 8. (b) thickness dependence of the apparent spin Hall conductivity $\sigma_{SH}$ of Bi$_{0.53}$Sb$_{0.47}$ [43]. Reprinted with permission from Chi *et al.*, Sci. Adv. 6, eaay2324 (2020). (c) Fermi level, (d) Resistivity $\rho_{xx}$ and 2D carrier density $|n_{2D}|$ for of (Bi$_{1-x}$Sb$_x$)$_2$Te$_3$ with different Sb percentages. (e) Switching current density $|J_c|$ and effective damping-like spin-orbit torque field vs the Sb ratio of (Bi$_{1-x}$Sb$_x$)$_2$Te$_3$ [105]. Reprinted with permission from Wu *et al.*, Phys. Rev. Lett. 123, 207205 (2019).

### 4. Magnetization switching of Van der Waals magnet
### 4.1 Van der Waals magnet

The recent discovery of van der Waals magnets (e.g., Cr$_2$Ge$_2$Te$_6$ [82], CrI$_3$ [83], etc.) has attached remarkable attention in the field of magnetism and spintronics. While the origin of the long-range magnetic order is still under debate, it has been suggested to have a close correlation with the suppression of thermal fluctuations by magnetic anisotropy. Note that in absence of magnetic anisotropy, no long-range magnetic order is expected by the Mermin–Wagner theorem [117] at finite temperature in a two-dimensional system. Van-der-Waals magnets provide a unique, highly tunable platform for spintronics. Most strikingly, the properties of van der Waals FMs, such as Curie temperature ($T_C$) [78,80], coercivity [79,80], and magnetic domain structure [81], can be tuned significantly by a variety of techniques (e.g., layer thickness, ionic liquid gating [78], proton doping [79], strain [80,81], exchange bias [118,119], interfacial proximity-effect [120], etc.). An interesting example is CrI$_3$, whose magnetic ordering depends on the number of layers and can be tuned by an external magnetic field. As shown in Fig. 10(a), the CrI$_3$ is ferromagnetic at 1 monolayer thickness, antiferromagnetic at 2 monolayer thickness, and ferromagnetic at 3 monolayer thickness. Ferromagnetic CrI$_3$ also shows a relatively square perpendicular magnetization loop [83].

Following the long-range ordering of magnetic lattices, magnetic materials can be grouped into ferromagnets, ferrimagnets, and antiferromagnets. In general, ferromagnets and ferrimagnets are considered more friendly than antiferromagnets to be integrated into electric circuits because their magnetization states can be electrically detected by anomalous Hall effect or tunnel magnetoresistance and efficiently switched by SOTs. In contrast, electrical detection and switching of collinear antiferromagnets [121-123] are generally much more challenging [124], despite the recent discovery of magnetoresistance and anomalous Hall in non-collinear antiferromagnets Mn$_3$Sn [34,125,126]. For this reason, spin-torque switching of magnetization is mostly studied and better understood in ferromagnetic and ferrimagnetic systems than in antiferromagnets. Our discussion below will be focused on van der Waals ferromagnets [116-134].

The van-der-Waals magnet CrBr$_3$ ($T_C$ = 34 K)[127,128] ($T_C$ = 34 K), CrI$_3$ ($T_C$ = 45 K) [83], Cr$_5$Te$_8$ [129] and VI$_3$ ($T_C$ = 60 K) [130] have perpendicular magnetic anisotropy but low Curie temperature. So far, room-temperature ferromagnetism and low-temperature perpendicular magnetic anisotropy have been reported for van der Waals materials FeTe [131], Fe$_4$GeTe$_2$ (Fig. 10(b))[132], Fe$_5$GeTe$_2$ [133], CrTe [134], CrTe$_2$ (Fig. 10(c))[135-137], Cr$_{1+\delta}$Te$_2$ [138], Cr$_2$Te$_3$ [139], Cr$_3$Te$_4$ [140]), CrSe [141], and Fe$_3$GaTe$_2$ (Fig. 10(d))[142]. In Fig. 11, we summarize the representative results of the Curie temperature and magnetization of relatively thin van der Waals magnets (note that $T_C$ of van der Waals magnets is strongly thickness dependent). While Fe$_n$GeTe$_2$ can have good PMA at low temperatures and Cr$_n$Te$_m$ and CrSe are relatively stable in air, they lose square hysteresis loops at room temperature [Fig. 10(b) and 10(c)]. The recently discovered Fe$_3$GaTe$_2$ [142] is an outstanding van der Waals ferromagnet that can have both a high Curie temperature ($T_C$ ≈ 350 - 380 K) and large PMA energy density ($K_u$ ≈ 4.8×10$^5$ J m$^{-3}$) [Fig. 10(d)]. Searching for Van der Waals magnets with room-temperature ferromagnetism, strong perpendicular magnetic anisotropy, and high stability in the air at the same is expected to be an active topic in the field.

### 4.2 Magnetization switching of van der Waals ferromagnets

Spin-orbit torque switching of van der Waals ferromagnets was first demonstrated in perpendicularly magnetized Fe$_3$GeTe$_2$/Pt bilayers [108,109], where the spin current generated by the SHE in the Pt exerts a damping-like spin torque on the Fe$_3$GeTe$_2$ [Fig. 12(a)]. Interestingly, despite the small layer thicknesses and small magnetization of the Fe$_3$GeTe$_2$, the Fe$_3$GeTe$_2$/Pt samples have a high depinning field (coercivity) and strong perpendicular magnetic anisotropy such that they typically require a large current density of $\sim 10^7$ A cm$^{-2}$ [108,109] as well as an in-plane magnetic field [Fig. 12(a)]. As indicated by the anomalous Hall resistance, the Fe$_3$GeTe$_2$ was also only partially switched, with the switching ratio of 20%-30% in [108] and 62% in [118], probably due to the non-uniformity of the magnetic domains with the van der Waals layer.

Cr$_2$Ge$_2$Te$_6$ is another well-studied van der Waals ferromagnet. Spin-orbit torque switching of Cr$_2$Ge$_2$Te$_6$ has been demonstrated in Ta/Cr$_2$Ge$_2$Te$_6$ bilayers (Curie temperature < 65 K) at a low current density of $5\times10^5$ A cm$^{-2}$ at 4 K, with an in-plane assisting magnetic field of 200 Oe [143]. Zhu et al. [110] reported SOT switching of Cr$_2$Ge$_2$Te$_6$/W with interface-enhanced Curie temperature of up to 150 K [Fig. 12(b)].

Current-induced magnetization switching has also been realized in all van der Waals heterojunctions. Nearly full magnetization switching (88%) has been reported in MBE-grown Cr$_2$Ge$_2$Te$_6$/(Bi$_{1-x}$Sb$_x$)$_2$Te$_3$ bilayers [144][Fig. 13(a)]. In the (Bi$_{0.7}$Sb$_{0.3}$)$_2$Te$_3$/Fe$_3$GeTe$_2$ bilayer [145], the threshold current density for the magnetization switching is $5.8\times10^6$ A cm$^{-2}$ at 100 K [Fig. 13(b)]. Field-free switching magnetization has been reported in the exfoliation-fabricated WTe$_2$/Fe$_{2.78}$GeTe$_2$ bilayers by a current along the low symmetry axis ($9.8\times10^6$ A cm$^{-2}$, $T$ =170K)[53]. In the same bilayer structure, Shin et al. also realized magnetization switching at a current density of $3.9\times10^6$ A cm$^{-2}$ ($T$ = 150 K, $H_x$ = 300 Oe, Fig. 13(c)) [146].

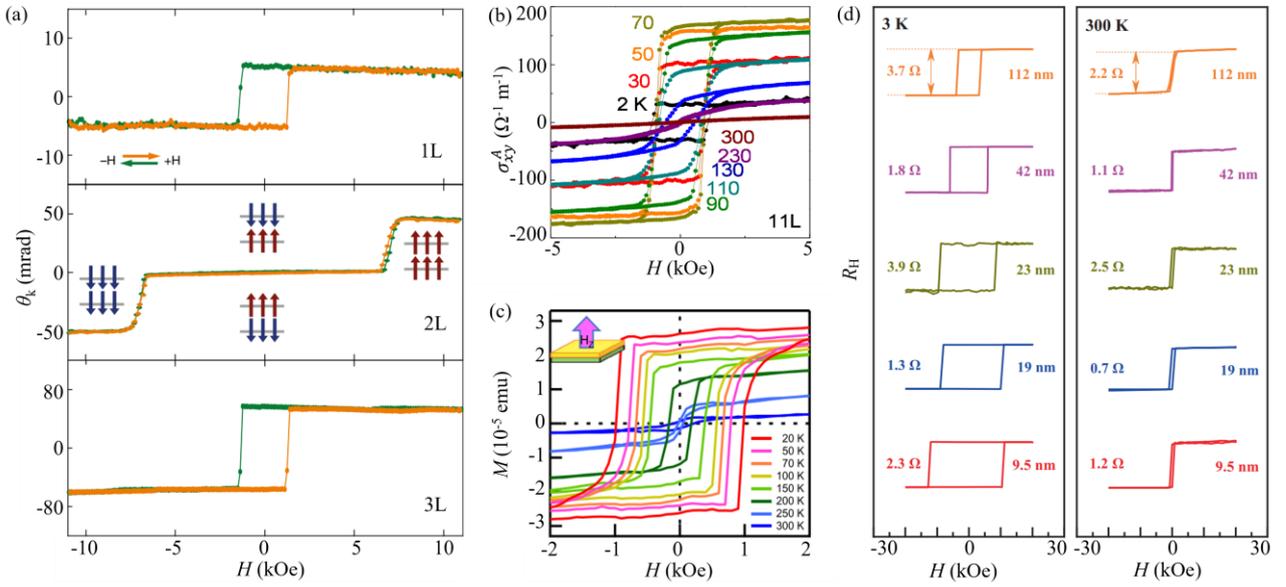

FIG. 10. (a) Kerr rotation vs perpendicular magnetic field for monolayer (1L), bilayer (2L), and trilayer (3L) CrI$_3$ flake [83]. Reprinted with permission from Huang et al., Nature 546, 270 (2017). (b) Hall conductivity hysteresis loop of a 11-monolayer-thick Fe$_4$GeTe$_2$ crystal at various temperatures [132]. Reprinted with permission from Seo et al., Sci. Adv. 6, eaay8912 (2019). (c) Out-of-plane magnetization hysteresis loop of 7 monolayer CrTe$_2$ at different temperatures along the out-of-plane direction [137]. Reprinted with permission from Zhang et al., Nat. Commun. 12, 2492 (2021). (d) Anomalous Hall resistance hysteresis (of Fe$_3$GaTe$_2$ with different thicknesses at 3 K and 300 K [142]. Reprinted with permission from Zhang et al., Nat. Commun. 13, 5067 (2022).

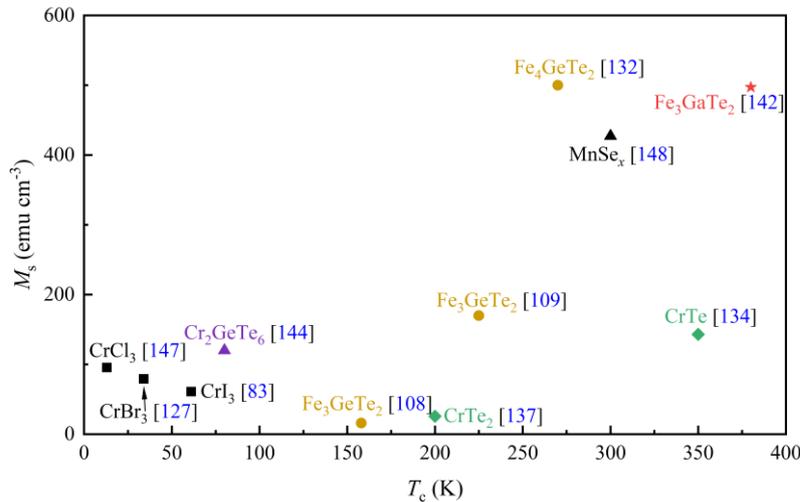

FIG. 11 Saturation magnetization $M_s$ vs Curie temperature $T_C$ of representative van der Waals ferromagnets.

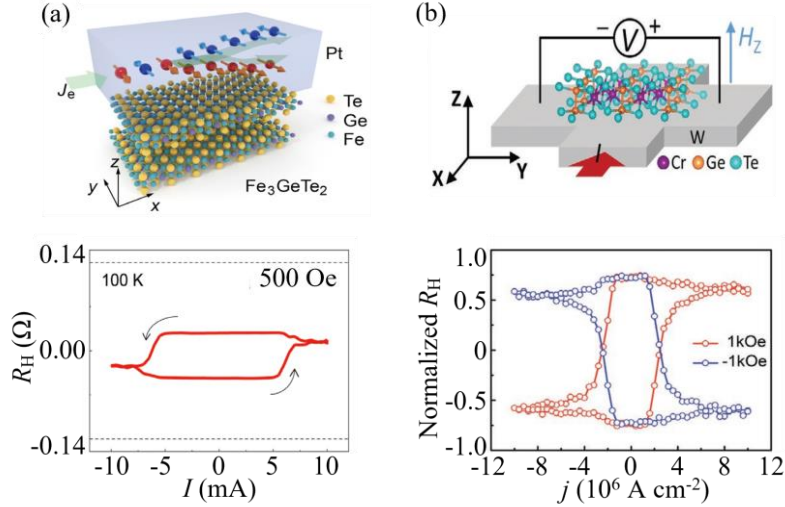

FIG. 12. (a) Current-driven perpendicular magnetization switching for Fe$_3$GeTe$_2$(4)/Pt(6) bilayer under an in-plane magnetic field of 500 Oe at 100 K [108]. Reprinted with permission from Wang *et al.*, Sci. Adv. 5, eaaw8904 (2019). (b) Current-driven perpendicular magnetization switching for Cr$_2$Ge$_2$Te$_6$(10)/W(7) bilayer under in-plane magnetic fields of ± 1 kOe at 150 K [110]. Reprinted with permission from Zhu *et al.*, Adv. Funct. Mater. 32, 2108953 (2022).

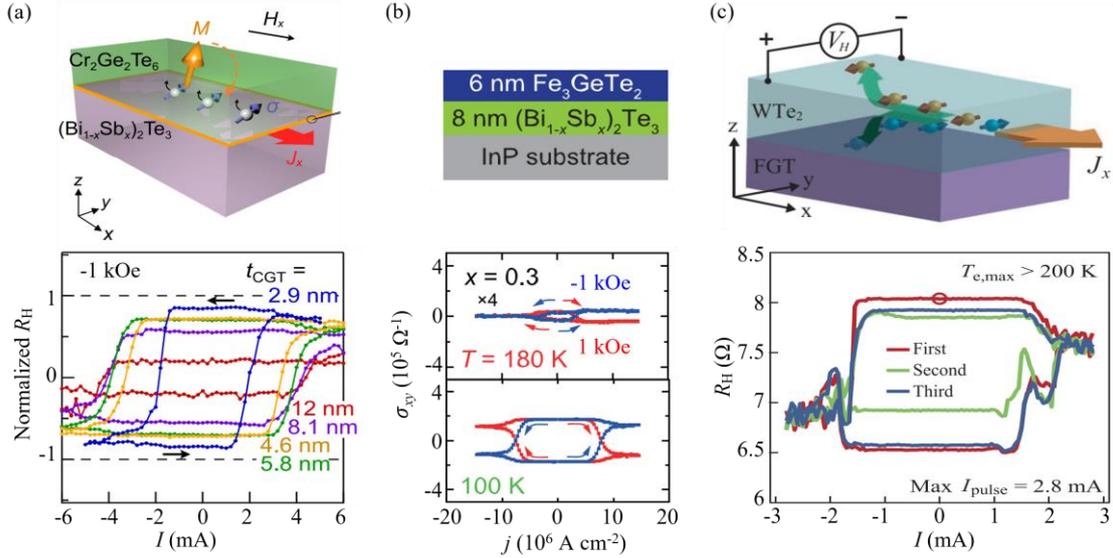

FIG. 13. Current induced switching of van der Waals magnets. (a) Normalized anomalous Hall resistance vs current density for (Bi$_{1-x}$Sb$_x$)$_2$Te$_3$(6)/Cr$_2$Ge$_2$Te$_6$(t) (*x* = 0.5) with different Cr$_2$Ge$_2$Te$_6$ thicknesses under an in-plane magnetic field of 1 kOe at 2 K [144]. Reprinted with permission from Mogi *et al.*, Nat. Commun. 12, 1404 (2021). (b) Anomalous Hall conductivity vs current density for (Bi$_{1-x}$Sb$_x$)$_2$Te$_3$(8)/Fe$_3$GeTe$_2$(6) at different temperatures under an in-plane magnetic field of 1 kOe [145]. Reprinted with permission from Fujimura *et al.*, Appl. Phys. Lett. 119, 032402 (2021). (c) Hall resistance vs current density for Fe$_3$GeTe$_2$(7.3)/WTe$_2$(12.6) under in-plane magnetic field $H_x$ = 300 Oe. The Hall resistance varies during three consecutive current scans due to temperature rise towards the Curie temperature [146]. Reprinted with permission from Shin *et al.*, Adv. Mater. 34, 2101730 (2022).

## 5. Simplifying models of switching current density

In this section, we provide a quantitative understanding of the switching current densities in the van der Waals system by considering the simplifying models. the transverse spin damping-like SOT efficiency per unit current density ($\xi_{DL}^j$) of a heterostructure with PMA inversely correlates to the critical switching current density ($j_c$) in the spin-current-generating layer via Eq. (4) in macrospin limit [149,150] and via Eq. (5) in the domain wall depinning limit [151-153], i.e.,

$$j_c \approx e\mu_0 M_s t_{FM} (H_k - \sqrt{2}|H_x|)/\hbar \xi_{DL}^j, \quad (4)$$

$$j_c = (4e/\pi\hbar) \mu_0 M_s t_{FM} H_c / \xi_{DL}^j, \quad (5)$$

where $e$ is the elementary charge, $\hbar$ is the reduced Planck constant, $\mu_0$ is the permeability of vacuum, $H_x$ is the applied field along the current direction, and $t_{FM}$, $M_s$, $H_k$, and $H_c$ are the thickness, the saturation magnetization, the effective perpendicular anisotropy field, and the perpendicular coercivity of the driven magnetic layer FM, respectively.

However, recent experiments [154] on heavy metal/magnet bilayers have shown that neither Eq. (4) nor Eq. (5) can provide a reliable prediction for the switching current and $\xi_{DL}^j$ and that there is no simple correlation between $\xi_{DL}^j$ and the critical switching current density of realistic perpendicularly magnetized spin-current generator/ferromagnet heterostructures. As shown in Table I, the same is true for the van der Waals systems. The macrospin analysis does not seem to apply to the switching dynamics of micrometer-scale samples so that the values of $\xi_{DL}^j$ determined using the switching current density and Eq. (4) can produce overestimates by up to hundreds of times ($\xi_{DL,macro}^j$ and $\xi_{DL,macro}^j/\xi_{DL}^j$ in Table I). A domain-wall depinning analysis [Eq. (5)] can either under- or over-estimated $\xi_{DL}^j$ by up to tens of times ($\xi_{DL,DW}^j$ and $\xi_{DL,DW}^j/\xi_{DL}^j$ in Table I). These observations consistently suggest that the switching current or "switching efficiency" of perpendicular heterostructures in the micrometer or sub-micrometer scales cannot provide a quantitative estimation of $\xi_{DL}^j$.

While the underlying mechanism of the failure of the simplifying models remains an open question, it is obvious that Joule heating during current switching of the resistive or low Curie-temperature van der Waals systems can have a rather significant influence on the apparent switching current density. As shown in Fig. 13(c), the anomalous Hall resistance hysteresis loop drifts for three consecutive current scans because of Joule heating [146].

TABLE I. Comparison of spin-torque efficiencies determined from the harmonic response or ST-FMR ($\xi_{DL}^j$) and magnetization switching ($\xi_{DL,DW}^j$, $\xi_{DL,macro}^j$) of PMA samples, which is calculated from Eq. (5) and Eq. (4) using the applied external magnetic field ($H_x$), saturation magnetization ($M_s$), the perpendicular coercivity ($H_c$), and the effective perpendicular anisotropy field ($H_k$) of the driven magnetic layer, and the critical magnetization switching current density ($j_c$). The value of $\xi_{DL,macro}^j$ is estimated to be negative for Te$_2$(10)/CoTb(6) [97] and ZrTe$_2$(7.2)/CrTe$_2$(1.8) [94] because according to the original reports an in-plane field greater that the effective perpendicular anisotropy field was applied.

| Sample | Technique | $M_s$ (emu cm$^{-3}$) | $H_c$ (Oe) | $H_k$ (kOe) | $H_x$ (Oe) | $j_c$ (MA cm$^{-2}$) | $\xi_{DL}^j$ | $\xi_{DL,DW}^j$ | $\xi_{DL,macro}^j$ | $\xi_{DL,DW}^j/\xi_{DL}^j$ | $\xi_{DL,macro}^j/\xi_{DL}^j$ |
|---|---|---|---|---|---|---|---|---|---|---|---|
| Te$_2$(10)/CoTb(6) [97] | sputtering | 48 | 60 | 0.33 | 900 | 0.7 | 0.2 | 0.47 | -5.8 | 2.4 | -29 |
| ZrTe$_2$(7.2)/CrTe$_2$(1.8) [94] | MBE | 100 | -- | 0.2 | 700 | 18 | 0.014 | -- | -0.12 | -- | -8.5 |
| Bi$_2$Se$_3$(7.4)/CoTb(4.6) [76] | MBE | 280 | 300 | -- | 1000 | 2.8 | 0.16 | 2.7 | -- | 17 | -- |
| Bi$_2$Se$_3$(6)/Gd$_x$(FeCo)$_{1-x}$(15) [77] | MBE | 46 | 160 | 0.35 | 1000 | 2.2 | 0.13 | 0.98 | 10 | 7.6 | 78 |
| Bi$_x$Se$_{(1-x)}$(4)/Ta(0.5)/ CoFeB (0.6)/Gd(1.2)/CoFeB(1.1)[113] | sputtering | 300 | 30 | 6 | 80 | 0.43 | 8.67 | 1.2 | 180 | 0.13 | 21 |
| (BiSb)$_2$Te$_3$(6QL)/Ti (2)/CoFeB(1.5) [105] | MBE | 868 | 30 | 2.24 | 100 | 0.12 | 2.5 | 6.3 | 345 | 2.5 | 138 |
| Bi$_2$Te$_3$/Ti(2) (6QL)/CoFeB(1.5) [105] | MBE | 868 | 27 | 2.06 | 100 | 2.4 | 0.08 | 0.287 | 16 | 3.5 | 194 |
| Bi$_2$Te$_3$(8)/MnTe(20) [112] | MBE | 175 | 100 | ≈50 | 400 | 6.6 | 10 | 1.0 | 397 | 0.10 | 40 |
| Bi$_{0.9}$Sb$_{0.1}$(5) /Mn$_{0.45}$Ga$_{0.55}$(3) [115] | MBE | 240 | 4500 | 10 | 3500 | 1.1 | 52 | 57 | 50 | 1.1 | 0.96 |
| SnTe(6QL)/Ti(2)/CoFeB(1.5) [105] | MBE | 868 | 53 | 2.18 | 100 | 1.5 | 1.41 | 0.91 | 27.5 | 0.65 | 20 |
| Fe$_3$GeTe$_2$(4)/Pt(6) [108] | exfoliation | 16 | 125 | 11 | 500 | 11.6 | 0.12 | 0.013 | 0.86 | 0.11 | 7.2 |
| Fe$_3$GeTe$_2$ (15)/Pt(5) [109] | exfoliation | 170 | 750 | 30 | 3000 | 20 | 0.14 | 1.8 | 50 | 13 | 355 |

## 6. Conclusion and outlook

We have reviewed recent advances in spin-orbit torque and resultant magnetization switching in van der Waals systems. Van der Waals materials provided unique opportunities for spintronics because of their diversity of materials, the flexibility of preparation (e.g., by mechanical exfoliation), and strong tunability by interface effects. Van der Waals TMDs such as WTe$_2$ also exhibit potential as a spin current source of both transverse spins and exotic perpendicular and longitudinal spins. Bismuth-based topological insulators and their sputter-deposited disordered counterparts are shown to generate giant damping-like SOT with the efficiency of up to 1-3 orders of magnitudes greater than 5$d$ metals. Moreover, van der Waals materials that are magnetic are also interesting for spin-orbitronics due to their highly tunable magnetism and low magnetization at small thicknesses since the damping-like and field-like effective SOT fields exerted on the magnetic layer by a given spin current scale inversely with the thickness and magnetization of the magnetic layer. Efficient switching of several van der Waals magnets by spin current has been demonstrated.

Despite these exciting progress, essential challenges have remained to overcome for spin-orbit torque switching of van der Waals systems:

(i) While the generation of different spin components has been demonstrated, the efficiencies are typically quite low. It has remained unclear as to whether and how the efficiency of generating exotic perpendicular and longitudinal spins by the low-symmetry TMDs can be improved significantly to be compelling for practical SOT applications.

(ii) Some Bi-based topological insulators and alloys have both giant effective spin Hall ratio and resistivity at

the same time, the latter is undesirable for metallic SOT devices that require energy efficiency, high endurance, and low impedance. It would be interesting if new uniform, stable spin-orbit materials can be developed to provide damping-like SOT efficiency of greater than 1 but substantially less resistive than the yet-known topological insulators.

(iii) So far, large-area growth of van der Waals magnets that have high Curie temperature, strong perpendicular magnetic anisotropy, and high stabilities against heating and atmosphere at the same time has remained a key obstacle that prevents van der Waals magnets from applications in spintronic technologies. Breakthroughs in the fabrication of such application-friendly van der Waals magnets are in urgent need.

(iv) While spin-orbit torque switching of magnetization has been demonstrated in Hall-bar devices containing van der Waals magnets, TMDs, or/and topological insulators, the simplifying models of macrospin rotation and domain wall depinning cannot provide an accurate prediction for the switching current density. So far, the quantitative understanding of the switching current remains a fundamental problem.

(v) From the point of view of magnetic memory and logic application, switching of thermally stable nanodots of van der Waals magnets, such as the free layers of nanopillars of magnetic tunnel junctions, by current pulses of picosecond and nanosecond durations. Efforts are also required on the demonstration and optimization of the key performance, e.g., the endurance, the write error rates, the retention, and the tunnel magnetoresistance.

**Acknowledgments**
This work was supported partly by the National Key Research and Development Program of China (Grant No. 2022YFA1204004) and by the National Natural Science Foundation of China (Grant No.12274405).